LETTER

# Scattering analysis of nano-scale chiral structures in optical vortex using Method of Moments

Chenxu Wang [1, a)], Hideki Kawaguchi[2], Hiroaki Nakamura[1,3], Koichi Matsuo[4], Masahiro Katoh[4]

**Abstract** This paper presents a full 3D Method of Moments (MoM) simulation for analysis of scattering from nano-scale chiral ribbon structures under optical vortex illumination to aim to investigate electromagnetic mechanism of circular dichroism (CD) and optical vortex dichroism (VD). Both twist and helical nano-ribbons are modeled as dispersive materials scatterer. CD and VD are evaluated quantitatively by differences in scattering power. Numerical results demonstrate that angular momentum of optical light induces dichroic response effectively.
**Classification:** Antennas and propagation

## 1. Introduction

Chirality is an essential characteristic of many biological molecules, such as proteins and DNA, which exhibit helical structures. Circular dichroism (CD) (see Fig.1), which measures the difference in scattering or transmission power for left and right circularly polarized light, has been widely employed to identify the chirality of the biomolecules [1]. Although the detailed electromagnetic mechanisms of CD have not yet been fully clarified, it is understood generally that CD is attributed to the interaction between the spin angular momentum (SAM) of circularly polarized light and chiral structures. Accordingly, in contrast to circularly polarized light, it is considered that optical vortex (OV) beams are employed for measurement of chirality of biomolecules (vortex dichroism: VD) [2], since the optical vortex carries orbital angular momentum (OAM) in addition to spin angular momentum (SAM) [3], [4]. Therefore, optical vortex carries much larger angular momentum than that of circularly polarized light, and it is expected that VD provides us more sensitive measurement of chirality than that of CD. However, detail electromagnetic mechanism of CD and VD is still unclear, such vortex dichroism (VD) remains largely unexplored.

This study aims to evaluate the interaction between optical vortex and chiral materials such as twist and helical nano-scale ribbons [5] by using 3D numerical simulations. Then, chiral molecules of nano-ribbons typically have sizes of ranging from a few nanometers to several tens of nanometers, while the wavelengths of optical light, which is sensitive for CD and VD, are approximately 780nm. Furthermore, the size of optical beam is typically several hundred times larger than wavelengths (see Fig.2). Accordingly, it is very difficult to use the finite-difference time-domain (FDTD) method for simulations of CD and VD due to the large difference between the size of the nanostructures and that of the incident optical beam. This paper investigates the scattering properties of chiral twist and helical nano-ribbons under 780nm optical vortex by using three-dimensional method of moments (3D MoM) in frequency domain.

## 2. Scattering analysis of chiral materials using MoM

### 2.1 Optical vortex of Laguerre-Gaussian mode

The Laguerre-Gaussian (LG) mode beams (1), which is a well-known solution to the paraxial approximation of the Helmholtz equation in cylindrical coordinates $(r, \emptyset, z)$, is one of typical optical vortex [3].

$$E(r,\emptyset,z) = \frac{C}{\left(1+\frac{z^2}{z_R^2}\right)^{\frac{1}{2}}}\left(\frac{\sqrt{2}r}{w(z)}\right)^l L_p^l\left(\frac{2r^2}{w^2(z)}\right)\exp\frac{-r^2}{w^2(z)} \cdot$$

$$\exp\frac{-ikr^2z}{2(z^2+z_R^2)}\exp(-il\emptyset) \cdot$$

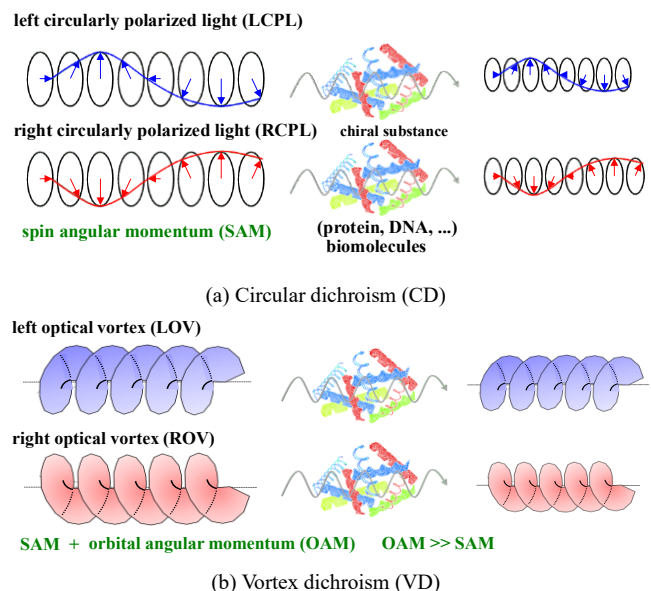

Fig.1 Overview of Circular dichroism (CD) and Vortex dichroism (VD)

[1] National Institute for Fusion Science, Toki 509-5292, Japan
[2] Muroran Institute of Technology, Muroran 050-0071, Japan
[3] Nagoya University, Nagoya 464-8601, Japan
[4] Hiroshima University, Hiroshima 739-0046, Japan
a) naka-lab@nifs.ac.jp

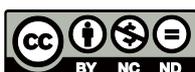

$$\exp\left(i(2p + l + 1)\tan^{-1}\frac{z}{z_R}\right) \quad (1)$$

where $L_p^l$ is Laguerre polynomial, $w(z)$ is beam radius, $z_R$ is Rayleigh range. In particular, the azimuthal index $l$ corresponds to the topological charge of the vortex beam, while the radial index $p$ determines the number of the radial modes. It is pointed out by Allen [3] that this L-G mode optical vortex beam carries spin angular momentum of $\sigma_z\hbar$ and orbital angular momentum of $l\hbar$ per one photon $\omega\hbar$, where $\sigma_z$ =+1, -1 and 0 are for left-handed, right-handed circular polarized light and linearly polarized light, respectively. In particular, $l$=0 corresponds to a plane wave, whereas $l \neq 0$ corresponds to an optical vortex with topological charge $l$. In Fig.2, examples of electric field vector distribution in $r - \emptyset$ vertical cross-section for L-G beam of $(\sigma_z, l)$=(a) (+1,0), (b) (-1,0), (c) (0, +1), (d) (0,-1), (e) (+1,+1), (f) (-1,-1) are depicted.

## 2.2 3D Method of Moments (MoM)

To treat extremely small size scatterer compared with incident field wavelength, we here employ 3D Method of Moments (MoM) in the frequency domain, which is based on the following magnetic field integral equation (MFIE) and electric field integral equations (EFIE) [6],

$$\boldsymbol{B}(\omega, \boldsymbol{x}) = \boldsymbol{B}_{ext}(\omega, \boldsymbol{x}) + \frac{1}{4\pi}\int_S e^{-i\frac{\omega}{c}|\boldsymbol{x}-\boldsymbol{x}'|}\left\{\left(\frac{(\boldsymbol{x}-\boldsymbol{x}')}{|\boldsymbol{x}-\boldsymbol{x}'|^3} + i\frac{\omega}{c}\frac{(\boldsymbol{x}-\boldsymbol{x}')}{|\boldsymbol{x}-\boldsymbol{x}'|^2}\right) \times (\boldsymbol{n}' \times \boldsymbol{B}(\omega, \boldsymbol{x}')) - \left(\frac{(\boldsymbol{x}-\boldsymbol{x}')}{|\boldsymbol{x}-\boldsymbol{x}'|^2} + i\frac{\omega}{c}\frac{(\boldsymbol{x}-\boldsymbol{x}')}{|\boldsymbol{x}-\boldsymbol{x}'|}\right)(\boldsymbol{n}' \cdot \boldsymbol{B}(\omega, \boldsymbol{x}')) + i\frac{\omega}{c^2}\frac{\boldsymbol{E}(\omega, \boldsymbol{x}')}{|\boldsymbol{x}-\boldsymbol{x}'|} \times \boldsymbol{n}'\right\} dS' \quad (2)$$

$$\boldsymbol{E}(\omega, \boldsymbol{x}) = \boldsymbol{E}_{ext}(\omega, \boldsymbol{x}) + \frac{1}{4\pi}\int_S e^{-i\frac{\omega}{c}|\boldsymbol{x}-\boldsymbol{x}'|}\left\{i\omega\boldsymbol{n}' \times \frac{\boldsymbol{B}(\boldsymbol{x}-\boldsymbol{x}')}{|\boldsymbol{x}-\boldsymbol{x}'|} - \left(\frac{(\boldsymbol{x}-\boldsymbol{x}')}{|\boldsymbol{x}-\boldsymbol{x}'|^3} + i\frac{\omega}{c}\frac{(\boldsymbol{x}-\boldsymbol{x}')}{|\boldsymbol{x}-\boldsymbol{x}'|^2}\right)(\boldsymbol{E}(\omega, \boldsymbol{x}') \cdot \boldsymbol{n}') - \left(\frac{(\boldsymbol{x}-\boldsymbol{x}')}{|\boldsymbol{x}-\boldsymbol{x}'|^3} + i\frac{\omega}{c}\frac{(\boldsymbol{x}-\boldsymbol{x}')}{|\boldsymbol{x}-\boldsymbol{x}'|^2}\right) \times (\boldsymbol{E}(\omega, \boldsymbol{x}') \times \boldsymbol{n}')\right\} dS' \quad (3)$$

where $\boldsymbol{x}$ and $\boldsymbol{x}'$ denote the observation and source points on the surface $S$, $\boldsymbol{n}'$ is the outward surface normal at $\boldsymbol{x}'$. Then, to consider dispersive material scatterer such as gold, silver, the EFIE is applied for inside region of the scatterer and the MFIE is for outside open region (vacuum) (Fig.3). These integral equations are discretized over the scatterer surface, and the resulting linear system (Fig.4) is solved to obtain unknown values of tangential components of magnetic and electric fields, $\boldsymbol{K}=\boldsymbol{n} \times \boldsymbol{B}$, $\boldsymbol{K}_m = \boldsymbol{E} \times \boldsymbol{n}$ on the surface $S$.

## 3. Numerical examples

In this work, we aim to evaluate CD and VD for chirality of nano-ribbons quantitatively by using the 3D MoM. Au nanoribbons are adopted based on experimentally demonstrated chiral Au nanostructures reported by Liu, Oda et al [5]. There are two types of nano-ribbon, twist and helical nano-ribbons (Fig.5). It is assumed that L-G mode incident field wavelength $\lambda$ is 780nm and the nano-ribbons are made by gold, which has complex permittivity $\hat{\varepsilon}_r = -23.0 + i1.44$ for 780nm wavelength field. The beam waist size is selected to be 10$\lambda$. The twist/helical nano-ribbon structures follow experimentally reported nanoscale chiral geometries. Each ribbon has a total physical length of 0.5$\lambda$ and is tilted by 45° relative to the propagation axis. Because optical vortices exhibit a phase singularity and vanishing field intensity at the center, the placement of nano-ribbons is critical. Under plane-wave illumination (Fig.6 (a)), the twist/helical nano-ribbons are placed at the beam center. Under vortex illumination (Fig.6 (b)), the ribbons are positioned off-axis where the electric field is non-zero and exhibits the characteristic azimuthal dependence required for OAM–chirality interaction. In Fig.7 and 8, distributions of tangential component of magnetic field on material surface of $(\sigma_z, l)$=(a) (+1, 0), (b) (0, +1), (c) (+1, +1) are depicted for twist and helical ribbon, respectively. The scattered fields of circularly polarized light are stronger than those of vortex light. Radiation pattern of $(\sigma_z, l)$=(+1, +1) for (a) twist ribbon and (b) helical ribbon are shown in Fig.9. It was found that

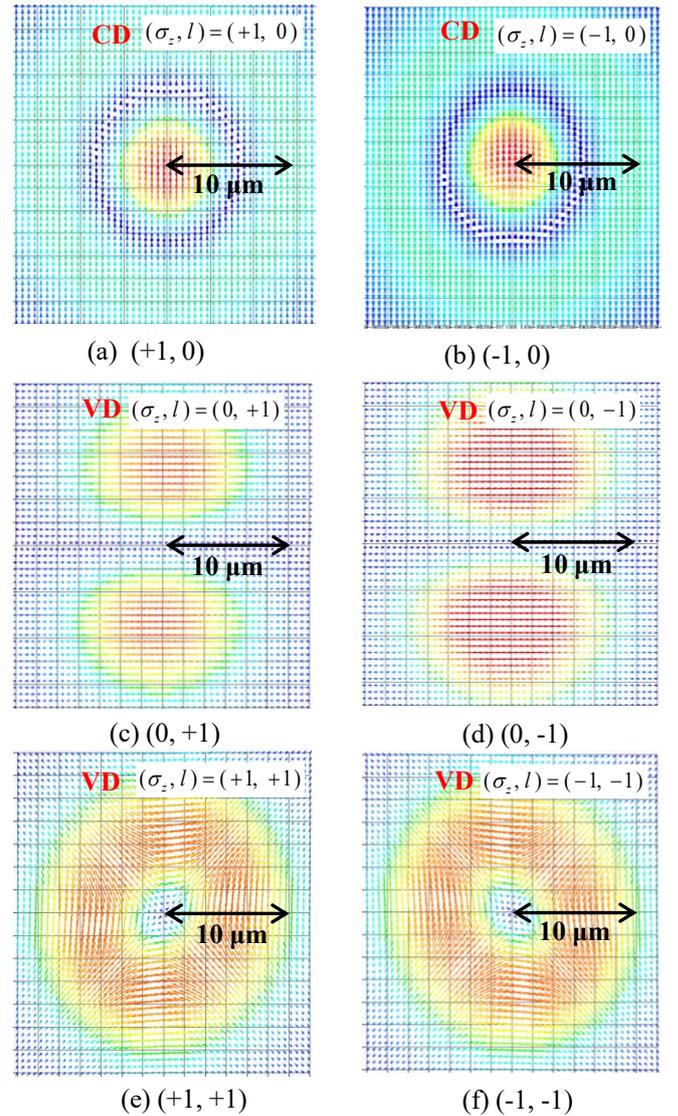

(a) (+1, 0)  (b) (-1, 0)

(c) (0, +1)  (d) (0, -1)

(e) (+1, +1)  (f) (-1, -1)

Fig.2 Examples of electric field vector distribution of L-G mode beam

the radiation pattern does not depend on the type of incident field and depends on only shape of scatterer.

The scattering power $P$ is computed by integrating the outward Poynting vector over a closed surface enclosing the ribbon. CD and VD are evaluated using the normalized asymmetry:

$$\text{CD/VD} = \frac{(P_+ - P_-)}{\frac{(P_+ + P_-)}{2}} \quad (4)$$

where $P_+$ and $P_-$ correspond to scattering under opposite SAM (LCPL and RCPL) or opposite OAM ($l$=+1, and $l$=-1).

### 3.2 CD and VD results

For the twist gold ribbon, the CD and VD reach approximately 30% for $(\sigma_z, l) = (\pm 1, 0)$ and $(\pm 1, \pm 1)$ incidences, while both vanish for $(0, \pm 1)$ (Table I), indicating that this structure exhibits almost no chiral asymmetry under purely OAM excitation. In contrast, the helical gold ribbon shows a much stronger chiral response. For the $(\pm 1, 0)$ and $(\pm 1, \pm 1)$ incidences, both CD and VD reach approximately 193%, even for the pure OAM cases $(0, \pm 1)$, small but noticeable asymmetries appear (Table II). These results demonstrate that the helical structure produces far stronger OAM induced chiral scattering than the twist ribbon.

### 4. Conclusion

In this study, the interaction between optical vortex carrying OAM and nano-scale chiral ribbon structures was investigated using 3D Method of Moments formulation. In contrast, a non-chiral ribbon does not exhibit any CD or VD response, confirming the role of structural chirality. By modeling twist and helical nano-ribbons with gold parameters, the scattering characteristics under both circularly polarized light and LG mode vortex illumination were evaluated. The numerical results revealed that, while CD arises from the coupling between the SAM of incident field and the chiral structures, VD exhibits a significantly larger response due to the additional OAM carried by the optical vortex.

Table I  CD/VD results for twist ribbon

| Scatterer twist gold | | | |
|---|---|---|---|
| Incident $(\sigma_z, l)$ | (+1, 0)(-1, 0) | (0, +1) (0, -1) | (+1, +1) (-1, -1) |
| CD/VD | 30.3% | 0% | 29.0% |

Table II  CD/VD results for helical ribbon

| Scatterer helical gold | | | |
|---|---|---|---|
| Incident $(\sigma_z, l)$ | (+1, 0)(-1, 0) | (0, +1) (0, -1) | (+1, +1) (-1, -1) |
| CD/VD | 193% | -10.5% | 193% |

### Acknowledgements

The computation was performed using Research Center for Computational Science, Okazaki, Japan (Project: 25-IMS-C100) and Plasma Simulator of NIFS. The research was supported by KAKENHI (Nos. 21H04456, 22K03572, 23K03362, 23K11190, 24K00613, 25H01640), by the NINS program of Promoting Research by Networking among Institutions (01422301) by the NIFS Collaborative Research Programs (NIFS24KIG002, NIFS25KIST062, NIFS25KIIT019).

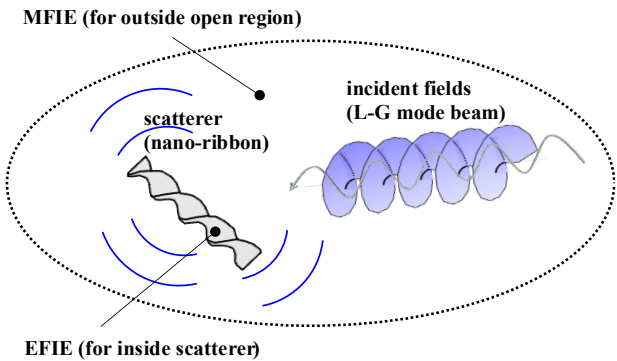

Fig.3 3D MoM analysis for volume model scatterer with dispersive materials

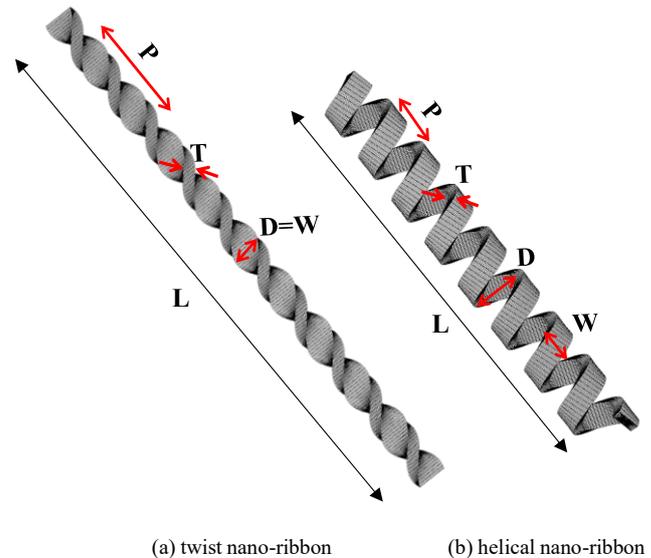

(a) twist nano-ribbon    (b) helical nano-ribbon
Fig. 5 Numerical model of nano-ribbons
(For the twist ribbon (a), the total length is L=390 nm, the pitch is P=78 nm, the ribbon width and rotation radius are W=D=20 nm, and the ribbon thickness T=10 nm. For the helical ribbon (b), the total length is L=390 nm, the pitch is P=48.75 nm, the ribbon width and rotation radius are W=D=22 nm, and the ribbon thickness T=10 nm.)

Fig.4 Configuration of linear equation of MoM for dispersive materials

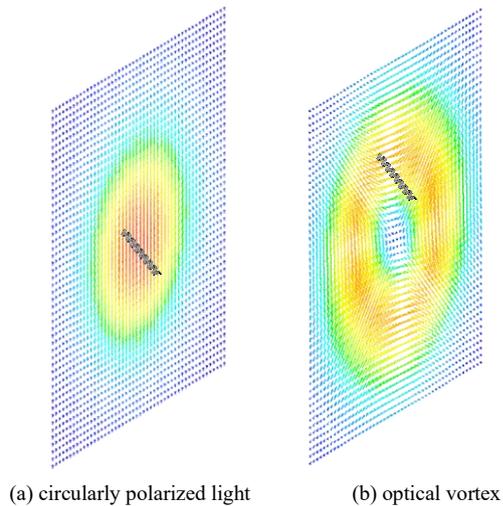

(a) circularly polarized light  (b) optical vortex

Fig.6 Position of scatterer in incident field

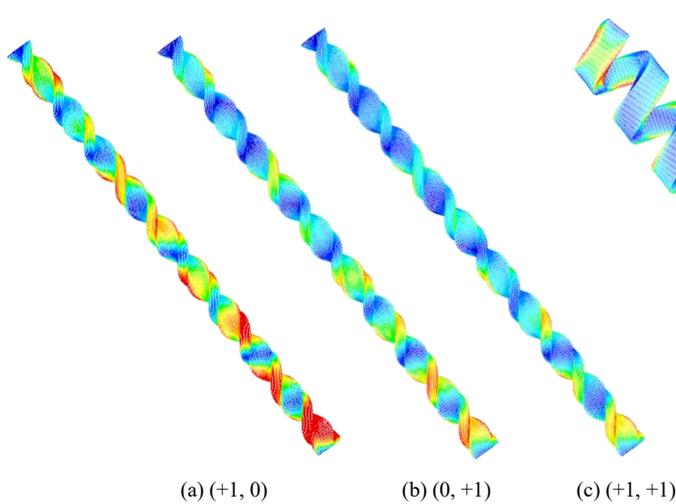

(a) (+1, 0)   (b) (0, +1)   (c) (+1, +1)

Fig.7 distribution of **n** × **B** for twist ribbon.

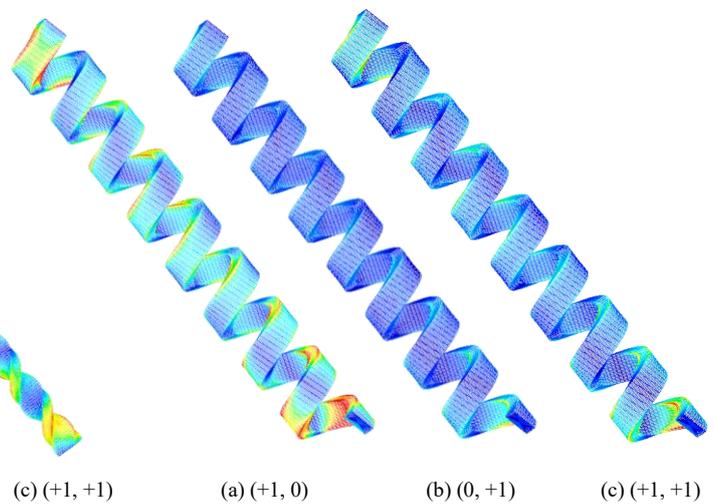

(a) (+1, 0)   (b) (0, +1)   (c) (+1, +1)

Fig.8 distribution of **n** × **B** for helical ribbon

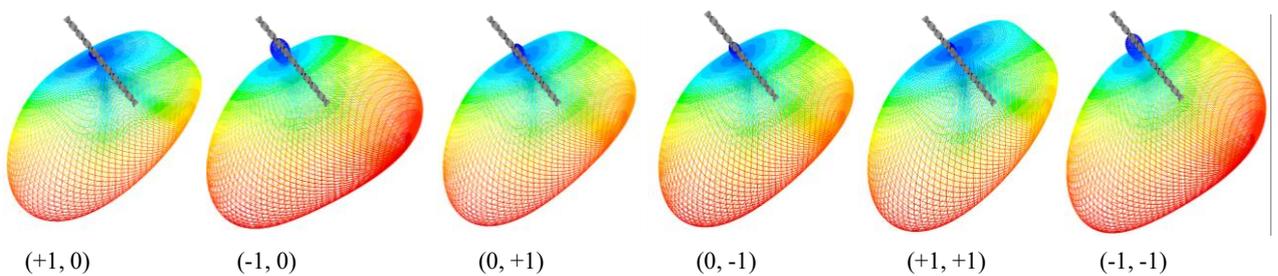

(+1, 0)   (-1, 0)   (0, +1)   (0, -1)   (+1, +1)   (-1, -1)

(a) twist ribbon

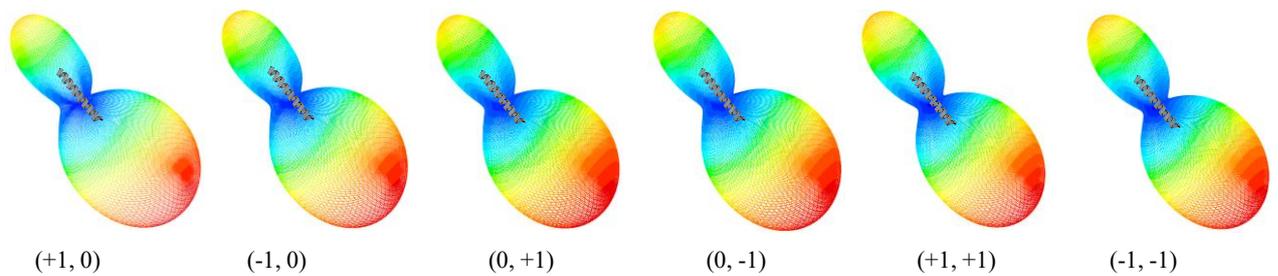

(+1, 0)   (-1, 0)   (0, +1)   (0, -1)   (+1, +1)   (-1, -1)

(b) helical ribbon

Fig.9 radiation pattern